\documentclass[iop,usenatbib]{emulateapj}

\def\1{\c{c}}
\def\2{\c{C}}
\def\3{\.{I}}
\def\4{\"{a}}
\def\5{{\i}}
\def\6{$\beta$}
\def\7{\"{o}}
\def\8{\"{O}}
\def\9{\c{s}}
\def\0{\c{S}}
\def\*{\"{u}}
\def\?{\"{U}}
\def\;{\u{g}}
\def\:{\u{G}}
\usepackage{color}

\slugcomment{The Astrophysical Journal Draft Version June 9, 2014}

\shorttitle{RRC in Gamma-ray Emitting MM SNR 3C 391}
\shortauthors{Ergin et al.}

\begin{document}

%%% Title of the Paper
\title{Recombining Plasma in the Gamma-ray Emitting Mixed-Morphology Supernova Remnant 3C 391}

%%% Author Names and Addresses
\author{T. Ergin*, A. Sezer}
\affil{TUBITAK Space Technologies Research Institute, ODTU Campus, 06531, Ankara, Turkey}
\affil{Bogazici University, Physics Department, Bebek, 34342, Istanbul, Turkey}
\email{* tulun.ergin@tubitak.gov.tr}
\author{L. Saha, P. Majumdar, A. Chatterjee}
\affil{Saha Institute of Nuclear Physics, Kolkata, West Bengal 700064, India}
\and
\author{A. Bay\5rl\5, E. N. Ercan}
\affil{Bogazici University, Physics Department, Bebek, 34342, Istanbul, Turkey}

%%% Abstract
\begin{abstract}
A group of middle-aged mixed-morphology (MM) supernova remnants (SNRs) interacting with molecular clouds (MC) has been discovered as strong GeV gamma-ray emitters by {\it Large Area Telescope} on board {\it Fermi Gamma Ray Space Telescope} (Fermi-LAT). The recent observations of the {\it Suzaku} X-ray satellite have revealed that some of these interacting gamma-ray emitting SNRs, such as IC443, W49B, W44, and G359.1-0.5, have overionized plasmas. 3C 391 (G31.9+0.0) is another Galactic MM SNR interacting with MC. It was observed in GeV gamma rays by Fermi-LAT as well as in the 0.3 $-$ 10.0 keV X-ray band by {\it Suzaku}. In this work, 3C 391 was detected in GeV gamma rays with a significance of $\sim$ 18 $\sigma$ and we showed that the GeV emission is point-like in nature. The GeV gamma-ray spectrum was shown to be best explained by the decay of neutral pions assuming that the protons follow a broken power-law distribution. We revealed radiative recombination structures of silicon and sulfur from 3C 391 using {\it Suzaku} data. In this paper we discuss the possible origin of this type of radiative plasma and hadronic gamma rays.
\end{abstract}

%%% Keywords
 
%%% Introduction
\section{Introduction}

About 10\% of the Galactic SNRs have radio shells and center-filled thermal X-rays, and they are called mixed-morphology SNRs \citep{rhopetre1998}. The first detected SNRs by Fermi-LAT were mostly middle-aged MM SNRs interacting with MC and exhibiting GeV gamma-ray luminosities distinctively higher than for the other detected SNRs, i.e. $\sim$ 10$^{35}$ $-$ 10$^{36}$ erg s$^{-1}$ for IC443, W28, W51C, W44, and W49B \citep{abdoIC4432010,abdoW282010,abdoW51C2009,abdoW442010,abdoW49B2010,castroslane2010}. Interactions of these MM SNRs with MC was shown by 1720 MHz OH masers \citep{yusefzadeh1995, frail1996, green1997, claussen1997, hewitt2009} and near-infrared observations \citep{keohane2007}. 

MM SNRs interacting with MC are interesting targets for the detection of gamma rays of hadronic origin, which provides clear evidence that these SNRs are sites of proton acceleration \citep{ackermann2013}. The hadronic mechanism is the production of two gamma rays from the decay of a neutral pion created in a proton-proton interaction during the passage of SNR shocks through the dense molecular material around the source. Therefore, the gamma-ray spectra of these SNRs rise steeply below 250 MeV  and at energies greater than 1 GeV they trace the parent proton energy distribution \citep{ackermann2013}. Interactions with MC may hint that MM SNRs are associated with star forming regions containing massive stars with strong stellar winds surrounded by circumstellar matter (CSM) and possibly these massive stars are the progenitors of MM-type remnants. When the supernova (SN) blast wave breaks out of the CSM into the ISM, its velocity rapidly rises and the particle acceleration increases. \citet{shimizu2012} calculated the amount of gamma-ray emission from an SNR blast wave breaking out of the CSM.

In young SNRs, the shocks create an X-ray emitting plasma, called ionizing plasma (IP), where the thermal energy of electrons (kT$_e$) is higher than the ionization energy (kT$_z$). This plasma gradually reaches collisional ionization equilibrium (CIE), where an equilibrium state between recombination and ionization is established (kT$_e$ = kT$_z$). The X-ray studies of {\it ASCA} on six MM SNRs \citep{kawasaki2002, kawasaki2005} revealed the existence of recombining plasma (RP) for IC443 and W49B, where kT$_z$ is higher than the kT$_e$. Recently, the X-ray Imaging Spectrometer, XIS, \citep{koyama2007} onboard {\it Suzaku} \citep{mitsuda2007} has discovered strong radiative recombination continuum (RRC) features from six MM SNRs, i.e. IC443, W49B, G359.1-0.5, W28, W44, and G346.6-0.2 \citep{yamaguchi2009, ozawa2009, ohnishi2011, sawada2012, uchida2012, yamaguchi2013}. There are two main scenarios describing the origin of the (electron cooling in) recombining (overionized) plasma in SNRs:

1) Thermal Conduction: When the hot ejecta inside the SNR interior, which is in the form of normal IP or CIE plasma, encounters cold MC, the electron energy will be transferred to the MC by thermal conduction and the electron temperature falls rapidly \citep{cox1999, shelton1999, orlando2008}. This condition then forms the RP. 

2) Adiabatic Cooling: If the CSM surrounding a progenitor is dense enough, CIE plasma will be formed at the early stages of the evolution of an SNR. When the blast wave breaks out of the dense CSM and expands rapidly into the rarefied ISM, the electron temperature drops due to the fast cooling by adiabatic expansion, which results in RP \citep{itohmasai1989, moriya2012, shimizu2012}.

Recently, \citet{lopez2013} produced overionization maps across the whole SNR to distinguish between the two RP scenarios in W49B. 

The Galactic SNR 3C 391 (G31.9+0.0), a member of the MM class, was suggested to be a result of an asymmetric core-collapse SN explosion of a massive ($\gtrsim$ 25M$_{\odot}$) progenitor star \citep{suchen2005}. The HI absorption measurements by \citet{radhakrishnan1972}, show that the distance to 3C 391 is at least 7.2 kpc (assuming a Galactocentric radius of 8.5 kpc) and for the emission without absorption indicate an upper limit of 11.4 kpc. 

In the radio band, 3C 391 is observed by VLA \citep{reynoldsmoffet1993} as a partial shell of 5$'$ radius with a breakout morphology, where the intensity of the radio emission in the shell rises in the bright northwest rim (NW) and drops and vanishes toward the southeast rim (SE). The CO(1-0) line observations of 3C 391 by \citet{wilner1998} showed that 3C 391 is embedded in the edge of an MC supporting the idea that the progenitor has exploded within the MC and that the SN blast wave has now broken out through the cloud boundary. Indirect evidence for 3C 391 expanding into a medium with different gas density comes from X-rays. In the X-ray band, 3C 391 was observed with {\it Einstein} \citep{wangseward1984}, {\it ROSAT} \citep{rhopetre1996}, {\it Chandra} \citep{chen2004}, and {\it ASCA} \citep{chenslane2001, kawasaki2005}. {\it ROSAT} and {\it Einstein} data revealed two bright X-ray peaks within the SNR: a brighter X-ray peak toward the interior of the weak SE radio rim and a fainter one in the interior of the bright NW radio shell. 

Using {\it ROSAT} observations, \cite{rhopetre1996} applied a single-temperature thermal model and obtained an absorbing column density of N$_{\rm{H}}$ $\sim$ 2.4 $\times$ 10$^{22}$ cm$^{-2}$ and electron temperature of kT$_{e}$ $\sim$ 0.5 keV. They also found enhanced abundances of Mg, Si, and S. \cite{chen2004} found that the X-ray spectra obtained from {\it Chandra} data can be best described by the VNEI model. The spectral fits showed that the diffuse emission have ionization parameters ($\tau$ = n$_e$t) close to or higher than 10$^{12}$ cm$^{-3}$ s. They concluded that the hot plasma in the SNR is very close to, or in, the ionization equilibrium. They found the electron temperature at $\sim$ 0.5 $-$ 0.6 keV and estimated an age of  $\sim$ 4 $\times$ 10$^3$ yr for the remnant. From the data of {\it ASCA} observation, \cite{kawasaki2005} found the electron temperature value as $\sim$ 0.53 keV by applying a non-equilibrium ionization (NEI) model to the spectra. They obtained an ionization timescale of $\tau \sim$ 2.5 $\times$ 10$^{12}$ cm$^{-3}$ s suggesting that the plasma has reached ionization equilibrium.

\citet{frail1996} observed two OH masers at 1720 MHz with velocities 105 and 110 km s$^{-1}$ coincident with the southeast and northeast rim of 3C 391, respectively, showing first clear evidence for 3C 391 interacting with an MC. The CO(1-0) data supports the evidence of SNR-MC interaction \citep{wilner1998}. Further evidence for shock interactions are obtained through the CS line observations by \citet{reachrho1999}, the measurements of strongly enhanced [O\,{\sc i}] 63 $\mu$m \citep{reachrho1996} at the NW rim of 3C 391, and the recent OH maser observations by \citet{hewitt2008}.

3C 391 was observed in GeV gamma rays by Fermi-LAT \citep{atwood2009} and it was listed in the 2nd Fermi-LAT  catalog \citep{nolan2012} as a point-source, called 2FGL J1849.3$-$0055. \citet{castroslane2010} analyzed the GeV data of 3C 391 and reported a $\sim$ 12 $\sigma$ detection. They showed that the peak of the significance map was shifted 4$'$ away from the NW edge of the radio shell. The spectrum of 3C 391 was best described as a power-law model with a spectral index of $\Gamma$ = $-$2.33 $\pm$ 0.11.  They found the integrated flux of 3C 391 as F(0.1$-$100 GeV) = (1.58 $\pm$ 0.26) $\times$ 10$^{-7}$ photons cm$^{-2}$ s$^{-1}$, \cite{castroslane2010}.  At TeV energies, H.E.S.S. reported integral flux upper limits at the 95\% CL in units of the flux of the Crab nebula as 0.8 Crab units, \citet{bochow2011}.

The gamma-ray emission from 3C 391 might be the result of hadronic interactions between the SNR shock and the associated MC. To understand if this is the case, we have performed a detailed modeling of the GeV gamma-ray spectrum. We have investigated the gamma-ray source morphology and variability. Moreover, we investigated the characteristics of the continuum radiation; thermal bremsstrahlung continuum or RRC by utilizing the superior spectral capabilities for diffuse sources of XIS onboard the {\it Suzaku}. We report on our results of RP in 3C 391 and discuss different scenarios of its origin. 

%%% Data Analysis And Results
\section{Data Analysis and Results} 
%%Gamma Rays
\subsection{ Gamma Rays} \label{gamma}
We analyzed the 3C 391 Fermi-LAT data taken in the period between 2008-08-04 and 2013-08-18. The events-data was taken from a circular region of interest (ROI) with a radius of 18$^{\circ}$ centered at the position of RA(J2000) = 18$^h$ 49$^m$ 26$^s\!$.40 and Dec(J2000) = $-$00$^{\circ}$ 55$'$ 37$''\!$.20 and the events suggested for Fermi-LAT Pass 7 for galactic point source analysis type were selected using \emph{gtselect} of Fermi Science Tools (FST). To prevent event contamination at the edge of the field of view caused by the bright gamma rays from the Earth's limb, we cut out the gamma rays with reconstructed zenith angles greater than 105$^{\circ}$.
%\clearpage
\begin{figure}[t]
\centering
\includegraphics[width=0.5\textwidth]{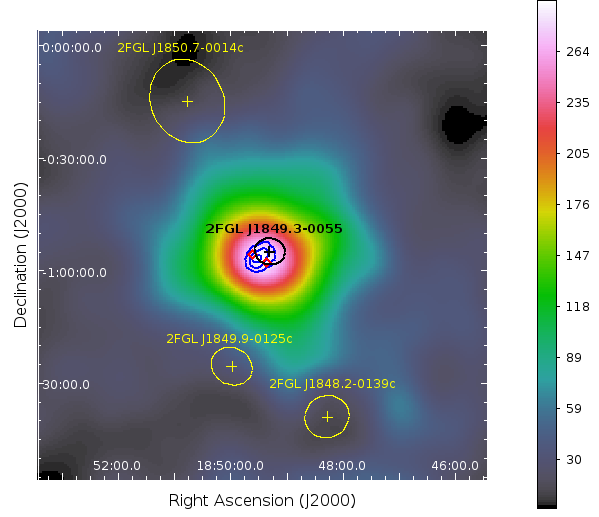}
\caption{\small{Gamma-ray TS map of the 3C 391 neighborhood with a bin size of 0$^{\circ}\!$.01 $\times$ 0$^{\circ}\!$.01. The blue contours show the {\it Suzaku} data from Figure \ref{figure_5a}, where 3 contours represent 14, 29, 43 cts. The yellow crosses and circles represent the 2nd Fermi-LAT catalog sources and the black cross and circle is the GeV source from the 2nd Fermi-LAT catalog corresponding to SNR 3C 391. Two red diamonds represent the two masers detected by \citet{frail1996}.}}
\label{figure_1}
\end{figure}  
For the rest of the analysis, we implemented \emph{pointlike} \citep{kerr2011, lande2012} (FST-v9r32p0) and the standard binned likelihood analysis tools (FST-v9r27p1), both based on the \emph{gtlike}, to cross-check the validity of the results. The analysis was performed within a square region of $\sim$ 25$^{\circ}\!$ $\times$ 25$^{\circ}\!$. The gamma-ray events in the data were binned in energy at 15 logarithmic steps between 250 MeV and 300 GeV. For the binned likelihood analysis \citep{abdo2009}, the matching energy dependent exposure maps were produced based on pointing direction, orientation, orbit location, and live-time accumulation of LAT. The point-spread function (PSF) of Fermi-LAT is up to 3$^{\circ}$ at 100 MeV and ~0$^{\circ}\!$.1 above 10 GeV. The large PSF of LAT means that at low energies, sources from outside the ROI can affect the analyzed source. To compensate for this and to ensure that the exposure map accounts for contributions from all the sources in the analysis region, exposure maps were created such that they included sources up to 10$^{\circ}$ outside the ROI. In addition, since at low energies the PSF is large, the exposure map should be expanded by another 10$^{\circ}$ to accommodate this additional exposure, \citet{abdo2009}. 

The spectral properties of the gamma-ray emission were studied by comparing the observation with models of possible sources in the ROI. Predictions were made by convolving the spatial distribution and spectrum of the source models with the instrument response function (IRF) and with the exposure of the observation. In the analysis we used the IRF version P7SOURCE$_{-}\!$V6. 

The model of the analysis region contains the diffuse background sources and all the point-like sources from the 2nd Fermi-LAT catalog located within a distance of 18$^{\circ}$ from the ROI center. We fixed all parameters of the point-like sources in the model, except 3 sources (shown in Figure \ref{figure_1} with yellow markers) within the distance of 2$^{\circ}$ from the best-fit location of 3C 391, where we set their normalization and spectral parameters free. The standard diffuse background model has two components: the diffuse Galactic emission (\emph{gal$_{-}$2yearp7v6$_{-}$v0.fits}) and the isotropic component (\emph{iso$_{-}$p7v6source.txt}), which is a sum of the extragalactic background, unresolved sources, and instrumental background. The normalization of the galactic diffuse background is set free during the analysis. The normalization of the isotropic component is fixed to one due to the difficulty to disentangle it from Galactic interstellar emission over limited regions.

The background and source modeling was done by the binned likelihood analysis using \emph{gtlike} of FST. To determine the best set of spectral parameters of the fit, we vary the parameters until the maximum likelihood is maximized. The detection of the source in this analysis is given by test statistics (TS) value, where larger TS values indicate that the null hypothesis (maximum likelihood value for a model without an additional source) is incorrect. This means that the detection significance is approximately equal to the square root of TS.

% Finding Best-fit Location and Spectrum
\subsubsection{Location and Spectrum}
Using both FST \emph{binned likelihood} (with FST-v9r27p1) and \emph{pointlike} (with FST-v9r32p0) analysis we detected 3C 391 with a significance of $\sim$ 18 $\sigma$. We computed the best-fit position within the ROI of 3C 391, which was found as longitude l = 31$^{\circ}\!$.879 $\pm$ 0$^{\circ}\!$.022 and latitude b = 0$^{\circ}\!$.022 $\pm$ 0$^{\circ}\!$.022. This best-fit position\footnote{RA(J2000) = 18$^h$ 49$^m$ 26$^s\!$.34 and Dec(J2000) = $-$00$^{\circ}$ 55$'$ 37$''\!$.35} enhanced the TS by 2.58 $\sigma$ over the position of 2FGL J1849.3$-$0055 in the 2nd Fermi-LAT catalog. Then the model was refitted using the best-fit position to compute the TS map (Section \ref{exte}) and the spectrum. 

\begin{figure}[t]
\centering
\includegraphics[width=0.5\textwidth]{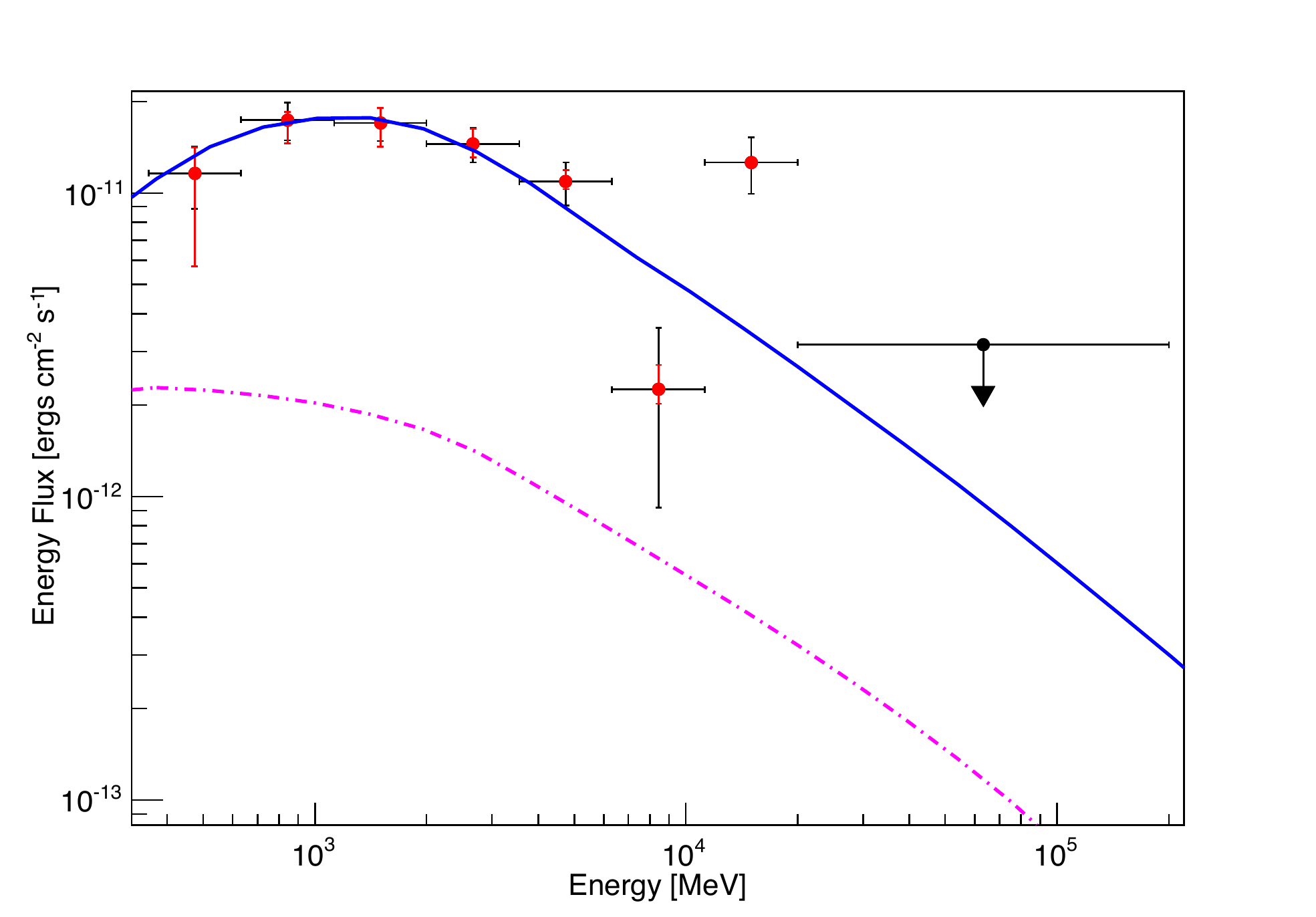}
\caption{\small{ The gamma-ray SED of 3C 391, where the Fermi-LAT spectral data points are represented with red filled circles and their corresponding statistical and systematic errors are shown in black and red lines, respectively. The thick blue line passing through the data points shows the hadronic model fit to the data. The dashed-dotted magenta line represents the bremsstrahlung spectrum. The parameters used to estimate the  emission spectra  both  for  hadronic and leptonic models are mentioned in Section \ref{modeling}. \\}}
\label{figure_4a}
\end{figure}

The observed spectral energy distribution (SED) of 3C 391 is shown in Figure~\ref{figure_4a}, where the data points are represented by red filled circles and their corresponding statistical and systematic errors are shown in black and red lines, respectively. To check the functional form of the spectrum, we first considered 3C 391 as a point-like source. First, the power-law (PL) function was fitted to the data between 250 MeV and 300 GeV, but we noticed that the spectrum deviates from a PL function. So, we checked, if the gamma-ray emission is better described by a log-parabola (LP) or a broken power-law (BPL) function, where the functional forms are as follows: 
\begin{itemize}
\item Log-parabola: \\
F(E)$^{LP}$ =  N$_{\circ}$ (E/E$_b$)$^{(\Gamma_1 + \Gamma_2 \mbox{{\small ln(E/E$_b$)}})}$
\item Broken Power-law:\\
 F(E)$^{BPL}$ =  N$_{\circ}$ (E/E$_b$)$^{-\Gamma_1}$ for  E $<$ E$_b$ \\
$~~~~~~~~~~~~~$ =  N$_{\circ}$ (E/E$_b$)$^{-\Gamma_2}$ for  E $>$ E$_b$ 
\end{itemize}

The results to these spectral fits are summarized in Table \ref{table_3}. Using different functions in fitting the spectrum of 3C 391, the likelihood ratio, TS, was used as a measure of the improvement of the likelihood fit with respect to the simple PL. The TS values of 337, 337, and 338 were found for PL, LP, and BPL fit, respectively. 

The PL resulted in spectral index of $\Gamma_1$ = 2.30 $\pm$ 0.03, which is in agreement with the best-fit power-law index value given for 3C 391 in the 2nd Fermi-LAT catalog ($\sim$ 2.19), \citet{nolan2012}. This result also matches to the results obtained by \citet{castroslane2010},  $\Gamma_1$ = 2.33 $\pm$ 0.11. Additionally, the LP fit results (shown in Table \ref{table_3}) were found to be in good agreement with the results in the 2nd Fermi-LAT catalog \citep{nolan2012}, which are $\Gamma_1$ = 2.35 $\pm$ 0.16 and $\Gamma_2$ = 0.308 $\pm$ 0.099 for a fixed E$_b$ value of 2430 MeV. 

The best-fit parameters for the BPL fit are N$_{\circ}$ = (1.15 $\pm$ 0.69) $\times$ 10$^{-11}$  MeV$^{-1}$ cm$^{-2}$ s$^{-1}$, $\Gamma_1$ = 1.28 $\pm$ 0.50, and $\Gamma_2$ =  2.50 $\pm$ 0.04, where the given uncertainties are statistical. The total energy flux was found as (6.28 $\pm$ 0.16) $\times$ 10$^{-11}$ erg  cm$^{-2}$ s$^{-1}$ with E$_b$ = 1060 $\pm$ 250 MeV.

\begin{table}
\scalebox{0.87}{
\begin{tabular}{|l|c|c|c|c|c|}
\hline Spectral                 & Photon Flux                & $\Gamma_1$   & $\Gamma_2$         & E$_b$       & TS       \\ 
           Model     & [ 10$^{-8}$ ph cm$^{-2}$ s$^{-1}$]  & $~$                     & $~$                           & [MeV]         & $~$     \\ \hline
           PL                            & 15.0 $\pm$ 1.7         & 2.30 $\pm$ 0.03          & $-$                            & $-$             & 337   \\ 
           LP                            & 7.14 $\pm$ 0.34       & 2.27 $\pm$ 0.04         & 0.15 $\pm$ 0.45     & 2430         & 337        \\ 
           BPL                         & 4.89 $\pm$ 0.57       & 1.28 $\pm$ 0.50         & 2.50 $\pm$ 0.04     & 1060          & 338      \\ \hline
\end{tabular}
}
\caption{Spectral fit parameters for PL, LP, and BPL between 250 MeV and 300 GeV assuming 3C 391 as a point-like source.}
\label{table_3}
\end{table}

Apart from the statistical uncertainties, there are systematic errors originating from the uncertainty of the Galactic diffuse background intensity. In order to calculate these systematic errors, we followed the prescriptions by \cite{abdoW51C2009} and \cite{castro2013} by varying the normalization value of the Galactic background by $\pm$6\% from the best-fit value and used these new frozen values of the normalization parameter to recalculate the SED of 3C 391. The systematic errors on the SED are shown in Figure \ref{figure_4a} in red color on top of the statistical errors.

%Extension
\subsubsection{Extension \label{exte}}
To investigate the morphology of 3C 391, we created a 2$^{\circ}$ $\times$ 2$^{\circ}$ TS map of 3C 391 and its neighborhood with a bin size of 0$^{\circ}\!$.01 $\times$ 0$^{\circ}\!$.01. The TS map shown in Figure \ref{figure_1} was produced with {\it pointlike} using a background model file, which contained all the point-like sources and diffuse sources, but excluded 3C 391 from the model. So, it shows the TS distribution of gamma rays originating dominantly from 3C 391. In Figure \ref{figure_1} the blue contours represent the {\it Suzaku} XIS image in the 0.3 $-$ 10.0 keV energy band (from Figure \ref{figure_5a}), the yellow crosses and circles represent the 2nd Fermi-LAT catalog sources and their statistical errors, respectively, and the black cross and circle represent the location and its statistical error of the GeV source from the 2nd Fermi-LAT catalog corresponding to 3C 391, respectively. The peak value of the gamma-ray significance coincides with the X-ray remnant. The red diamonds indicate the locations of the two OH masers reported by \citet{frail1996}. 

To search for the energy dependent morphology, we split the data set into two energy ranges (250 MeV $-$ 1 GeV and 1 $-$ 300 GeV) and computed the TS maps for each energy range. We found no significant gamma-ray excess at the location of 3C 391 for the energy range between 250 MeV  and 1 GeV, but 3C 391 was detected in the higher-energy range of 1 $-$ 300 GeV with a significance of $\sim$ 15 $\sigma$ using a BPL spectral model.

Additionally, using {\it pointlike} we have checked the extension of 3C 391 by fitting a disk template and assuming 3C 391 has a PL/BPL type spectrum. To detect the extension of a source, we use the TS of the extension (TS$_{ext}$) parameter, which is the likelihood ratio comparing the likelihood for being a point-like source (L$_{pt}$) to a likelihood for an existing extension (L$_{ext}$), TS$_{ext}$ = 2log(L$_{ext}$/L$_{pt}$). {\it pointlike} calculates TS$_{ext}$ by fitting a source first with a disk template and then as a point-like source. 

According to the extension studies by \cite{lande2012}, the extended source detection threshold is TS$_{ext}$ = 16, where the threshold is defined as the source flux at which the value of TS$_{ext}$ averaged over many statistical realizations is 16. From simulation studies, it is found that to resolve a disk-like extension of r = 0$^{\circ}\!$.1 at the level of TS$_{ext}$ = 16, the source must have a minimum flux of 3 $\times$ 10$^{-7}$ ph cm$^{-2}$ s$^{-1}$ for a spectral index value of 2.0 and a flux of 2 $\times$ 10$^{-6}$ ph cm$^{-2}$ s$^{-1}$ for a spectral index value of 2.5 \citep{lande2012}. So, for 3C 391 TS$_{ext}$ was found as 0.008 and 0.52 after a disk template fitting for 3C 391 having a PL and BPL type spectrum, respectively (Table \ref{table_5}). 
Both of the TS$_{ext}$ values are smaller than 16, which indicate that a disk-like extension with r $\sim$ 0$^{\circ}\!$.1 could not be resolved at the integrated flux level and spectral index values of 1.5 $\times$ 10$^{-7}$  ph cm$^{-2}$ s$^{-1}$ and 2.3 for the PL type spectrum and 4.9 $\times$ 10$^{-8}$  ph cm$^{-2}$ s$^{-1}$ and 2.5 ($\Gamma_{2}$) for the BPL type spectrum of 3C 391.
\begin{table}[h]
\begin{tabular}{|l|c|c|c|c|}
\hline Disk                              & Longitude                    & Latitude                       & Sigma                               & TS$_{ext}$ \\
           Model                           & [$^{\circ}$]                   & [$^{\circ}$]                   & [$^{\circ}$]                        & $~$               \\ \hline
           PL                                 & 31.87 $\pm$ 0.02      & 0.023 $\pm$ 0.017    & 0.10 $\pm$ 0.15             & 0.008            \\
           BPL                              & 31.88 $\pm$ 0.02	      & 0.028 $\pm$ 0.017    & 0.10 $\pm$ 0.26             & 0.52               \\ \hline  
\end{tabular}
\caption{$\!\!\!$Fit results of disk-like extension model applied to 3C 391 gamma-ray data between 250 MeV and 300 GeV for the PL and BPL type spectral models.}
\label{table_5}
\end{table}

%Variability and Pulsation
\subsubsection{Variability and Pulsation}
\begin{figure}[t]
\centering
\includegraphics[width=0.5\textwidth]{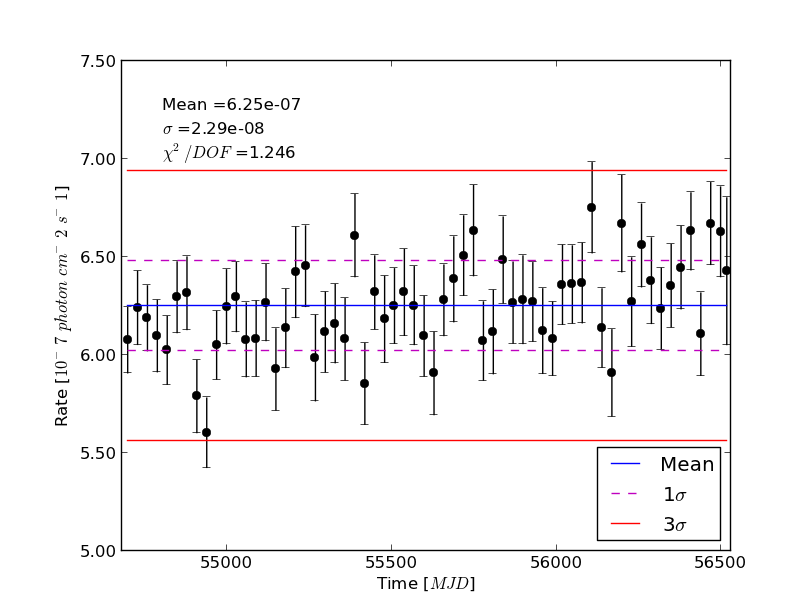}
\caption{\small{The monthly gamma-ray variability for 3C 391 with BPL-fit spectrum in the energy range of 0.25 $-$ 300 GeV}. \vspace{0.1cm}}
\label{figure_2}
\end{figure}

Variability or pulsations can effect the analysis results for 3C 391. So, we checked for both of these effects in the data. We searched for long term variability in the light curve of 3C 391 produced using the data from the circular region of 1$^{\circ}$ around the best-fit position. Figure \ref{figure_2} shows the 1-month binned light curve after fitting the spectrum with a BPL, where each flux point remains within 1 $\sigma$ and 3 $\sigma$. Fitting these flux points to a straight line (shown as a blue line in Figure \ref{figure_2}), yields a $\chi^2$/degrees of freedom (dof) of $\sim$ 1.25. Thus, we conclude that there is no long term variability observed in the close neighborhood of 3C 391. 

We have also checked if the spectral shape of 3C 391 fits to the standard spectrum of a pulsar, Power Law with Exponential Cutoff (PLEC). The best-fit cutoff energy is found as 28.80 $\pm$ 6.73 GeV, which is an order of magnitude away from the range of typical pulsar cutoff energies \citep{abdopulsar2010}. The PLEC fit didn't show a significant improvement over the PL, BPL, and LP spectral fits. 

%Comparison of Gamma-ray CO Data
\subsubsection{Molecular Environment}
\begin{figure}[t]
\centering
\includegraphics[width=0.5\textwidth]{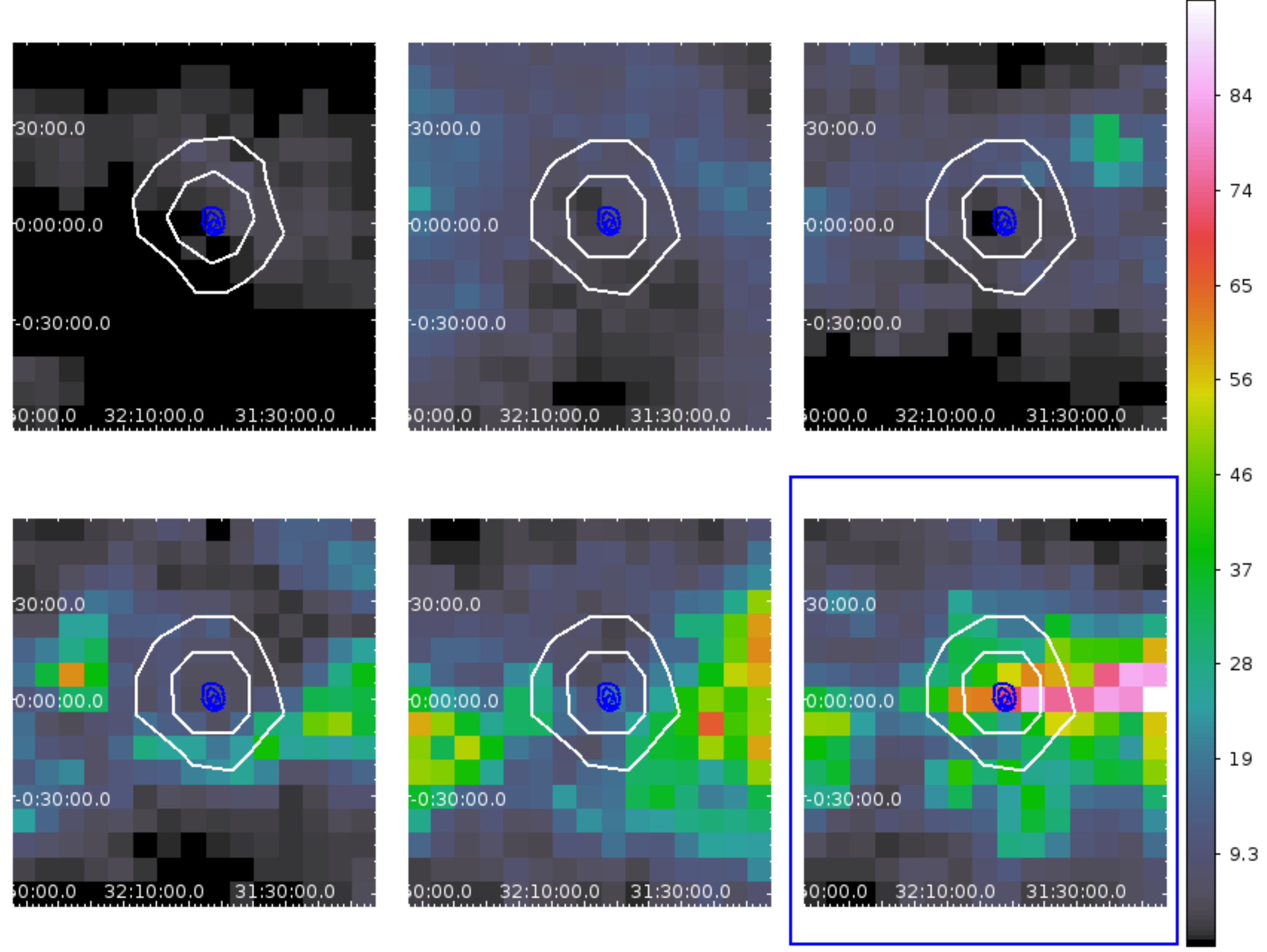}
\caption{\small{Maps in galactic coordinates (longitude-x-axis and latitude-y-axis) of the integrated CO intensity (W$_{CO}$) for 6 different velocity ranges (Top from left: [$-$50, 0], [0, 15], [15, 35]; Bottom from left: [35, 60], [60, 90], [90, 120] km s$^{-1}$) at the location of 3C 391 and its vicinity. For all maps, the range for W$_{CO}$ is fixed between 0 and 92.8 K km s$^{-1}$.} \vspace{0.25cm}}
\label{figure_3}
\end{figure}

To estimate the average density of the vicinity of 3C 391, we used the CO data of Harvard-Smithsonian Center for Astrophysics 1.2 m Millimeter-Wave Telescope \citep{dame1987}. We analyzed the CO gas in the whole velocity range integrated from $-$50 to 120 km s$^{-1}$, where the velocity intervals were divided such that each range included at least one cloud cluster peaking in temperature at a certain velocity.  Figure \ref{figure_3} shows the CO maps produced at different velocity ranges of [$-$50, 0], [0, 15], [15, 35] km s$^{-1}$ starting from top-left; and [35, 60], [60, 90], [90, 120] km s$^{-1}$ starting from bottom-left. The white contours represent the TS values of 3C 391 gamma-ray data at 41 and 83, and the blue contours are the X-ray counts at 14, 29, and 43.  

The velocity integrated CO intensity (W$_{CO}$) for the whole velocity range and for the region covering the whole X-ray remnant was found to be $\sim$ 110 K km s$^{-1}$. Since the CO sky maps are binned as 0$^{\circ}\!$.125 $\times$ 0$^{\circ}\!$.125, the area corresponding to the total W$_{CO}$ emission is (0$^{\circ}\!$.125)$^2$. We calculated W$_{CO}$ for each above mentioned velocity range and found that the highest contribution at the SNR's X-ray location came from the velocity range of  90 $-$ 120 km s$^{-1}$, which is also apparent in the CO sky maps in the blue framed panel on the bottom right corner of Figure \ref{figure_3}. 

Yet when we calculated the integrated CO intensity over the velocity and extent of the cloud, S$_{CO}$, we took all the velocity ranges into account: S$_{CO}$ = $\sum_i{(W_{CO}~A)_i}$, where $i$ represents the different velocity ranges. So, we found S$_{CO}$ =  1.72 K km s$^{-1}$ deg$^2$ for CO data used from the whole velocity range. For the dominant velocity range of 90 $-$ 120 km s$^{-1}$, we found S$_{CO}$ = 1.11 K km s$^{-1}$ deg$^2$.
Assuming a linear relationship between the velocity integrated CO intensity, W$_{CO}$, and the molecular hydrogen column density, N(H$_2$): 
\begin{eqnarray}
{N(H_2) \over W_{CO}} = (1.8 \pm 0.3) \times 10^{20}~\mbox{cm}^{-2} \mbox{K}^{-1} \mbox{km}^{-1} \mbox{s}^{-1}
\label{eqn:xfactor}
\end{eqnarray}
as given in \citet{dame2001}. Equation \ref{eqn:xfactor} gives 
\begin{eqnarray}
{M_{CO} \over  M_{\odot}}~=~1200~S_{CO}~d_{kpc}
\label{eqn:massratio}
\end{eqnarray}

$\!\!\!\!\!\!$where $d_{kpc}$ is the distance to the cloud in kpc. We calculated the total hydrogen column density as N(H$_2$) = 1.98 $\times$ 10$^{22}$ cm$^{-2}$ using the CO data in the whole velocity range. For the velocity range of 90 $-$ 120 km s$^{-1}$ we obtained N(H$_2$) = 1.28 $\times$ 10$^{22}$ cm$^{-2}$. 
The total mass of the clouds with velocities in the range of 90 $-$ 120 km s$^{-1}$ was found from Equation \ref{eqn:massratio} as  M$_{CO}$ = 6.9 $\times$ 10$^{4}$  M$_{\odot}$ using lower limit on distance to the cloud ($\sim$ 7.2 kpc). We estimated the size of the emission region as 15.7 pc. By assuming a spherical geometry of the cloud we computed the density of the source region to be 4.25 M$_{\odot}$ pc$^{-3}$ and the average density of protons to be 269 protons cm$^{-3}$. But using the upper limit of the distance to the cloud ($\sim$ 11.4 kpc) we obtained an upper limit to the average proton density,  671 protons cm$^{-3}$. Considering only the highest cloud velocity range, we recalculated the proton density as 173 and 435 protons cm$^{-3}$ for the source distances of 7.2 and 11.4 kpc, respectively. Averaging the proton density over all different combinations of the distance and velocity parameters we obtained 387 protons cm$^{-3}$. \citet{wilner1998} found $\sim$ 300 protons cm$^{-3}$, a value typical for giant MCs.

%Interpretation of gamma-ray results
\subsubsection{Modeling and Interpretation \label{modeling}} 

To determine if the observed gamma-ray SED of 3C 391 can be fitted by a hadronic model, we fit the gamma-ray spectrum resulting from the decay of neutral pions, $\pi^{\circ}$, following \citet{kelner2006}. In order to calculate the gamma-ray spectrum, we considered that the relativistic protons follow a BPL type spectrum:
 
 \begin{eqnarray}
{dN \over dE_p}& = & N_{1} E_p^{-\alpha} ~\mbox{for} ~E_p <E_{br} \nonumber \\
& = & N_{2} E_p^{-\beta} exp{\left(-{E_p \over E_{p_{max}}} \right )} ~ \mbox{for}~ E_{br}\leq E_p \leq E_{p_{max}}~~~~~
\label{eqn:pi0}
\end{eqnarray}
 
In Equation \ref{eqn:pi0}, E$_p$ is the proton energy and E$_{br}$ is the spectral break energy, where the spectral index changes from $\alpha$ to $\beta$. E$_{p_{max}}$ is the maximum energy of protons and during the fitting procedure we assumed it to be at 10 TeV. N$_1$ and N$_2$ are normalization constants. 

The best-fit parameters for the proton spectrum were obtained by a $\chi^2$-fitting procedure to the flux points.  The estimated parameters are $\alpha~ = ~2.48 ~\pm~ 0.45$, $\beta ~ = ~3.0 ~\pm~ 0.22$, and $E_{br}$ = 12 GeV. The $\chi^2$/dof is estimated to be $\simeq$ 1.6. The best-fit gamma-ray spectrum resulting from the decay of neutral pions for an ambient gas density of $\sim$ 387 cm$^{-3}$ is shown in Figure \ref{figure_4a} with the blue solid line. The estimated total energy can be written as $W_{p}  \simeq 5.81 \times 10^{48} \left({387 ~\mbox{cm}^{-3} \over n_H}\right)$ erg, where $n_H$ is the effective gas number density for p-p collision.  In addition to BPL spectrum, different proton spectra, like PL, LP, and PLEC were considered to explain the gamma-ray spectrum. However, we didn't find any significant difference in the estimated best-fit parameters for all the input proton spectra.  

To check the hadronic scenario from the energy point of view, we considered the energy from the SN explosion converted into accelerated protons, W$_p$ = L $\times$ $\tau_p$, where L is the gamma-ray luminosity and $\tau_p$ is the characteristic cooling time of protons. When the gamma-ray luminosity is dominated by hadronic emission, then $\tau_p$ =  5.3 x 10$^7$ (n/(1 proton cm$^{-3}$))$^{-1}$ yr \citep{aharonian2004}. Using the average proton density of 387 protons/cm$^3$ for n, we found $\tau_p$ = 1.37 $\times$ 10$^5$ yr. So, taking W$_p$ $\sim$ 5.81 $\times$ 10$^{48}$ erg, the luminosity of 3C 391 was found as L = 1.34 $\times$ 10$^{36}$ erg s$^{-1}$. 
 
\begin{figure}[t]
\centering
\includegraphics[width=0.52\textwidth]{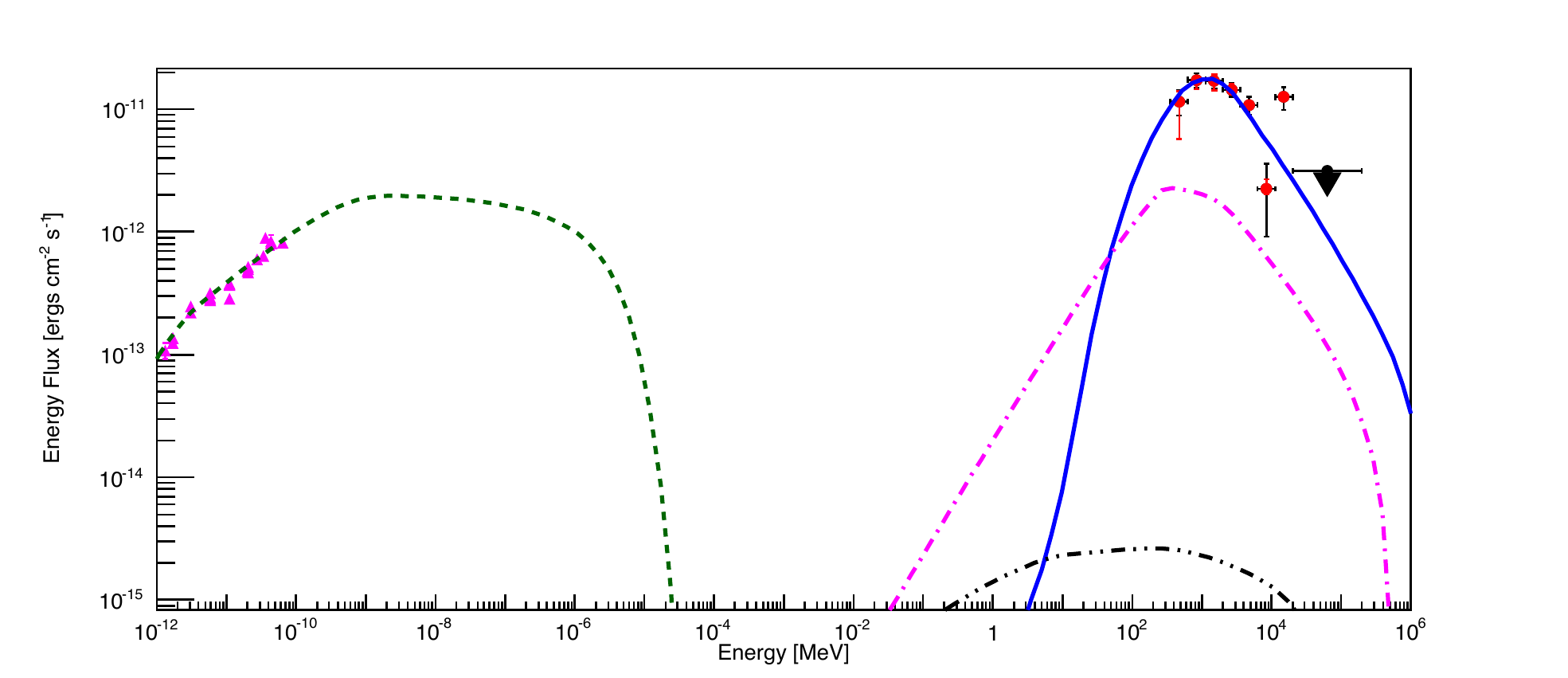}
\caption{{\small The models fit to the radio \citep{kassim1989} (magenta filled triangles) and gamma-ray data (red filled circles with their corresponding statistical and systematic errors) are shown with solid blue ($\pi^0\!$-decay spectral model component), green dashed (synchrotron emission), magenta dotted-dashed (non-thermal bremsstrahlung component), and black double-dot-dashed (IC emission component) lines.} \vspace{0.3cm}}
\label{figure_4b}
\end{figure}

We also calculated the contribution from the leptonic emission models \citep{blumenthalgould1970}. We found that the relativistic electrons can not account for the gamma-ray spectrum at GeV energies through inverse Compton (IC) and bremsstrahlung processes. We assumed the broken power-law type spectrum for electrons, which is similar to that considered for protons (Equation \ref{eqn:pi0}). The fit to the radio data \citep{kassim1989} gave a synchrotron spectral index $\sigma \simeq$ 0.55 ($S_\nu ~\propto~\nu^{-\sigma}$). Therefore, we took $\alpha~=~2.1$ in the electron spectrum before the break, because this parameter determines the shape of the synchrotron spectrum at radio wavelengths. On the other hand, $\beta$ can be found out from the fit to the observed gamma-ray spectrum. We considered an electron to proton ratio of 0.01 following the observed spectra of the Galactic cosmic electrons and protons. We then considered the magnetic field and the number of electrons in the emission volume such that synchrotron spectrum of electrons could explain the observed radio data as shown in Figure \ref{figure_4b}. To estimate the spectrum from leptonic models we used the following parameters: $\alpha$ = 2.1, $\beta$ = 3.0, $E_{br}$ = 7 GeV, B = 210 $\mu$G, n = 387 $\rm{cm}^{-3}$. We found the total energy of electrons as $W_e= 1.4 \times 10^{47}$ erg. Assuming that the gamma rays at GeV energies are produced by the same population of electrons, we estimated the IC spectrum by taking the cosmic microwave background radiation and interstellar background radiation fields following \cite{porter2008}. We found that neither the IC nor the bremsstrahlung emission could account for the observed gamma-ray fluxes shown in Figure \ref{figure_4b}.

 %%X-rays
\subsection{X-Rays} \label{x-rays}
%Observation and Data Reduction
\subsubsection{Observation and Data Reduction}
3C 391 was observed with XIS on board {\it Suzaku} on 2010 October 22, under the observation ID of 505007010 and an exposure time of $\sim$ 99.4 ks. Detailed descriptions of the {\it Suzaku} satellite, the XIS instrument, and the X-ray telescope are given in \citet{mitsuda2007}, \citet{koyama2007}, and \citet{serlemitsos2007}, respectively. The XIS system consists of four CCD cameras (XIS 0, 1, 2, and 3). One of the cameras (XIS1) uses a back-illuminated (BI) CCD while the others (XIS0, 2, and 3) use front-illuminated (FI) CCDs. XIS2 has not been functional due to an unexpected anomaly in 2006 November. The XIS was operated in the normal full-frame clocking mode.

For the data reduction we used {\it HEASoft} package version 6.11.1. The calibration database (CALDB: 20130305) was used and fitting was carried out in the X-ray spectral fitting package (\emph{xspec}) version 11.3.2 \citep{arnaud1996}. The redistribution matrix files (RMFs) of the XIS were produced by \emph{xisrmfgen} and \emph{auxillary} response files (ARFs) were generated by \emph{xissimarfgen} \citep{ishisaki2007}.
\begin{figure}[t]
\centering
\includegraphics[width=0.5\textwidth]{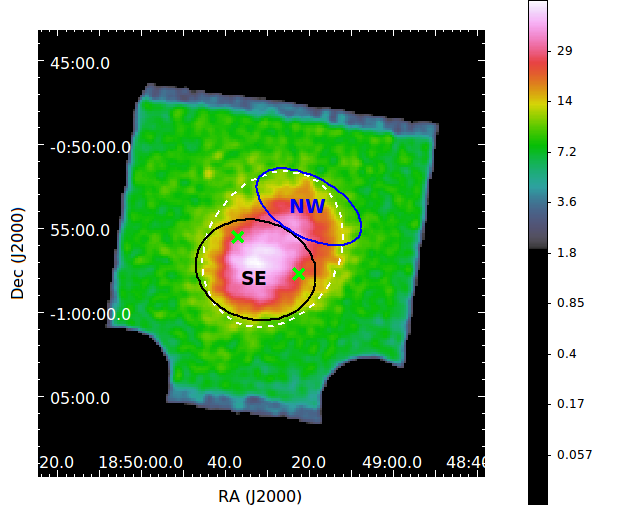}
\caption{\small{{\it Suzaku} XIS1 image of 3C 391 in the 0.3$-$10.0 keV energy band. The whole SNR region shown in white dashed ellipse and the NW and SE regions of the SNR shown in blue and black solid ellipses, respectively, are chosen for the spectral analysis. The two maser spots are represented with green crosses \citep{frail1996}. The bottom corners irradiated by the calibration sources of $^{55}$Fe are excluded.} \vspace{0.1cm}}
\label{figure_5a}
\end{figure}

%Background Estimation and Model Fitting
\subsubsection{Background Estimation and Model Fitting}
For the spectral analysis of 3C 391 we selected three regions, which are the whole SNR and the NW and SE regions of the SNR. These regions are shown on the XIS1 image of 3C 391 in the 0.3 $-$ 10.0 keV energy band in Figure \ref{figure_5a} as white dashed, blue solid, and black solid ellipses, respectively.  The region representing the whole remnant has a size of 4$'\!$.85 $\times$ 3$'\!$.94 centered at RA(J2000) = 18$^{\rm h}$ 49$^{\rm m}$ 28$^{\rm s}\!$.6, Dec(J2000) = $-$0$^{\circ}$ 56$'$ 16$''\!$.4. The reasons for this selection are given in the Discussion section.

The background for 3C 391 is a combination of the non-X-ray background (NXB), the Cosmic X-ray background (CXB) emission and the Galactic ridge X-ray emission, GRXE \citep{koyama1986}. First, we estimated the NXB data from night-Earth observations using the tool \emph{xisnxbgen} \citep{tawa2008}, and subtracted the NXB from the spectrum. We selected background region, the nearby blank sky region (Obs.ID 500009020) on the Galactic plane, consisting of the GRXE and the CXB. NXB-subtracted background spectrum was subtracted from the source spectrum using \emph{xspec}. The spectrum was binned to a minimum of 20 counts per bin using \emph{grppha} to allow use of the $\chi^2$ statistic.
\begin{figure}[t]
\centering
\includegraphics[width=0.47\textwidth]{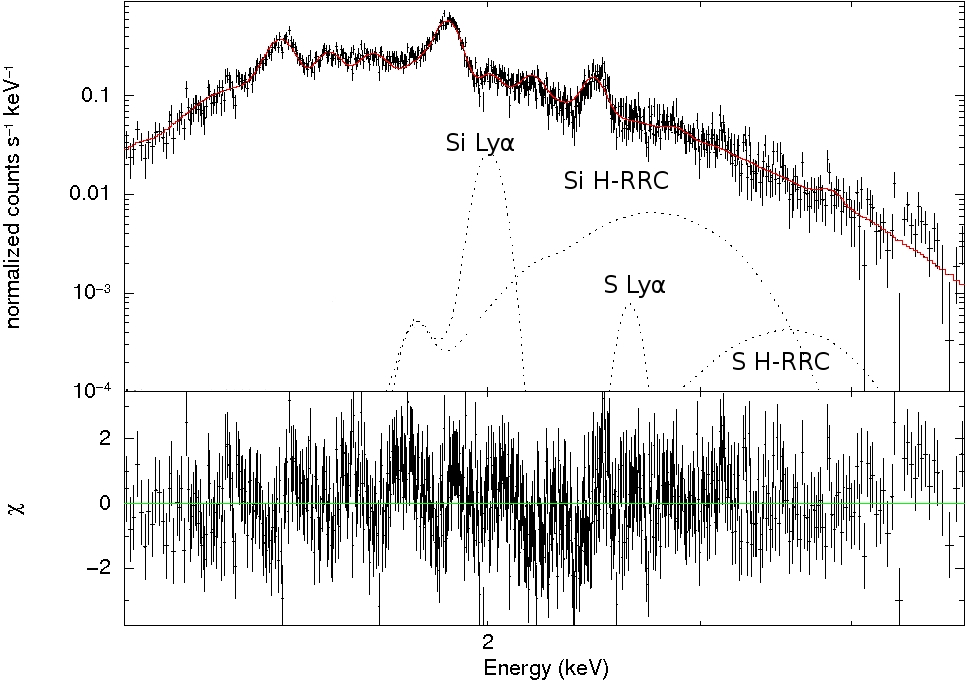}
\caption{\small{Background-subtracted FI spectrum of 3C 391 in the 1.0 $-$ 5.0 keV energy band fitted with an absorbed VNEI model with RRC and Ly$\alpha$ lines of the Si and S for the whole SNR. At the bottom of this panel, the residuals from the best-fit model are shown.} $\vspace{0.3cm}$}
\label{figure_5b}
\end{figure}

We first started the XIS analysis with the whole SNR region. We applied an absorbed (wabs in \emph{xspec}; \cite{morrison1983}) VNEI model for a NEI collisional plasma with variable abundances \citep{borkowski2001}, which gave the reduced $\chi^2$ value of 950.2/659 = 1.44 for the energy range of 1.0 $-$ 5.0 keV. During model fitting N$_{\rm{H}}$, kT$_e$, n$_e$t, and the abundances of Mg, Si, and S were free parameters, while the other elemental abundances were fixed to their solar values \citep{anders1989}. Residuals of the VNEI spectral fit show that there is a clear residual emission at energies of $\sim$ 2.0 keV and $\sim$ 2.6 keV. Therefore, we added two Gaussian components (gauss model in {\it xspec}) to the VNEI model. These two lines in the spectrum correspond to the H-like (Ly$\alpha$) lines of Si and S, which are the indicators of highly ionized plasma. We note that, we found Al K-shell emission at $\sim$ 1.58 keV from this remnant, as it was also found for G344.7$-$0.1 \citep{yamaguchi2012}, G350.1$-$0.3, and G349.7$+$0.2 \citep{yasumi2014}. We follow the prescription described in \citet{ozawa2009} for W49B to understand if the X-ray continuum comes from the thermal bremsstrahlung process or from RRC. We added the RRC model of H-like Si (2.666 keV) and S (3.482 keV). The added Ly$\alpha$ lines and the RRC components improved the quality of the fit ($\chi^2$/dof = 860/717). This suggest that these residuals are caused by the RRC of Si and S. Figure \ref{figure_5b} shows the background-subtracted FI spectrum fitted with the absorbed VNEI plus RRC models and the Ly$\alpha$ lines of the Si and S, in the energy range of 1.0 $-$ 5.0 keV. The same analysis steps described above were also applied for the NW and SE regions of 3C 391. The parameters (90\% confidence level) computed for the best-fit model obtained for the whole SNR and for the NW and SE regions are presented in Table \ref{table_2}.

%%% Discussion 
\section{Discussion}
\label{section:discussion}
%For Gamma rays
In this paper, we showed that the gamma-ray spectrum of 3C 391 can be described by a hadronic emission model, a clear evidence for acceleration of protons in this SNR. The neutral pion decay model assumed the protons to follow a BPL distribution with $\alpha$ index standing for the acceleration of cosmic rays in the shock and $\beta$ index represents the energy, above which protons escape from the SNR shell. 
Since, the bremsstrahlung spectrum depends linearly on the number density of ambient matter, for 3C 391 it can only account for the observed spectrum at GeV energies, if $n=3000~\rm{cm}^{-3}$. However, the spectrum has to be much steeper at $\sim$ 1 GeV to explain the observed spectrum at this energy. Introducing an abrupt break at $\sim$ 800 MeV in the electron spectrum can explain the observed fluxes at $\sim$ 1 GeV, but this break in the electron spectrum will make the synchrotron radio spectrum inefficient to explain the observed radio fluxes. In the case of IC emission process, total energy has to be $\sim 10^{51}$ erg to account for the observed spectrum. This means that almost all the energy released during the SN explosion has been transferred to the relativistic electrons, which is very unlikely. Moreover, the magnetic field needs to be $\leq$ 1 $\mu$G to explain the radio data. Additionally, density of the ambient matter has to be $\sim$ 0.3 cm$^{-3}$ to reduce the bremsstrahlung component, which will be inconsistent with our measured value of 387 cm$^{-3}$. 

The total gamma-ray luminosity was found as L = 1.34 $\times$ 10$^{36}$ erg s$^{-1}$, similar to the first GeV-emitting SNRs that were discovered by Fermi-LAT, e.g. IC443 \citep{abdoIC4432010}, W51C \citep{abdoW51C2009}, W44 \citep{abdoW442010}, and W49B \citep{abdoW49B2010}, all of which are MC interacting MM SNRs with gamma-ray luminosities higher than 10$^{35}$ erg s$^{-1}$. 

%Models for hadronic emission from SNRs
There are mainly two scenarios describing how hadronic gamma rays are produced in SNRs. The `crushed cloud' scenario describes the hadronic gamma rays as a product of interactions between the MC, compressed and shocked by the passage of the blast-wave of the SNR, and the relativistic protons inside the shocked MC. The relativistic protons can be either reaccelerated cosmic rays or freshly accelerated protons entering the radiatively compressed MC. In this scenario since the crushed clouds are thin, multi-GeV particles can escape from the shocked MC, which might be the reason of seeing a break in the proton spectrum \citep{blandfordcowie1982, uchiyama2010}. In the `escaped cosmic rays' scenario, escaping relativistic protons reach a nearby unshocked MC and produce $\pi^{\circ}\!$-decay gamma rays. For this scenario to happen, there must be GeV/TeV sources found outside the radio shell of 3C 391 that could produce these escaping protons \citep{gabici2009}. But there are no nearby cosmic ray sources to 3C 391 and all other sources were taken into account in the gamma-ray background model. 

The reason why the spectral break of protons is at $\sim$ 12 GeV for 3C 391 could be that relativistic particles are escaping from their acceleration sites, the shell of the SNR or the crushed MC, when the shell is expanding into the rarefied ISM during the earlier epochs of the SNR. Since particles at very high energies ($\sim$ TeV) can only be confined during the early stages of the SNR evolution, and because 3C 391 is a middle-aged SNR, most of the very high energy particles must have already escaped from the shell \citep{uchiyama2010}. The interactions between the SNR and MC, which are evident by the OH masers \citep{frail1996}, can enhance the particle escape. Alternatively, the SNR shock expanding in dense medium can be slowed down by the dense MC shifting the maximum particle energy to the GeV region \citep{ohira2011}. Assuming the BPL spectrum of protons without any spectral cutoff, we estimate that the differential flux of gamma rays at 1 TeV is $\sim$ 0.06\% of Crab nebula flux. TeV observations of this source with the upcoming Cherenkov Telescope Array (CTA) may provide more robust constraints of the various parameters of the input proton spectrum. 

%For X-rays
The massive progenitor star of 3C 391 inside an MC exploded and the shock waves expanded in the dense MC breaking out into a more rarefied ISM in the SW region, where most of the very high energy particles escaped causing a break in the proton spectrum. This `break-out' scenario could also be an explanation for observing recombining plasma in 3C 391. Using {\it Suzaku} data we discovered RRC of H-like Si and S at  $\sim$ 2.7 and $\sim$ 3.5 keV from the spectrum. 

We chose three regions to do X-ray spectral analysis, to compare the kT$_{e}$ values in different regions of 3C 391: the whole SNR, the NW region, and the SE region (Figure \ref{figure_5a}). We chose the NW and SE regions to check if there is any temperature gradient across the SNR. Finding a temperature gradient would be sign for electron cooling through thermal conduction mechanism. The NW region is closer to the site of denser molecular material, where the molecular density drops gradually toward the SE of 3C 391 \citep{wilner1998}. This is a good region to test the thermal conduction scenario of the RP. The SE region includes two OH maser spots; an indication that the SNR shell is breaking out of the MC and into the rarefied ISM \citep{frail1996}. By choosing this region, we aim for checking the adiabatic cooling scenario over ionized plasma. When we compare the kT$_e$ values of these regions, we found kT$_e$ for NW and SE region to be $\sim$ 0.61 keV and $\sim$ 0.54 keV, respectively. Since these values are very close to each other it is not possible to determine which cooling mechanism is dominating. 

Analyzing X-rays from the whole remnant and assuming a distance of  7.2 kpc for 3C 391 \citep{radhakrishnan1972}, we calculated the electron density $n_{\rm e}$ as $\sim$ 0.82 cm$^{-3}$ from the emission measure ($EM = n_{\rm e}n_{\rm H}V$, where $n_{\rm H}$ is the hydrogen density and $V$ is the volume of the X-ray emitting plasma). Then, from the relation $\tau$/$n_{\rm e}$, we found the age of 3C 391 as $\sim$ 69,000 yr using the best-fit $\tau$ value ($\sim$ 1.8 $\times$ $10^{12}$ cm$^{-3}$ s) of VNEI plus RRC model, higher than the SNR age found by {\it Chandra} \citep{chen2004}. 

The possible origin of the RP found in NW region might be due to the hot electrons getting in contact with cooler and denser MC impeding the expansion of the SNR shell. In the SE region of 3C 391, it is possible that the RP formed when SN blast wave expanded into the rarefied ISM and caused the electron temperature to drop through the adiabatic cooling mechanism. Both of these cooling mechanisms might have worked together in different regions of the SNR to produce RRC in 3C 391. To understand which scenario dominates in which part of 3C 391, detailed over-ionization maps need to be produced, as it is done for W49B \citep{lopez2013}. 

%%% Conclusion 
\section{Conclusion}
%For gamma rays
We have analyzed GeV gamma rays from 3C 391 and the spectrum revealed that the emission is most likely hadronic in origin. The gamma-ray spectrum follows the spectrum of parent protons, a BPL distribution with spectral index parameters of $\alpha$ = 2.48 and $\beta$ = 3.0 and a spectral break at $\sim$ 12 GeV. This suggests that protons are accelerated to high energies, possibly at the region of SNR shell breaking through the shocked MC into the rarefied ISM. The low breaking energy can be explained either by higher energy protons being already escaped through the thin crushed MC or the SNR shell at the earlier evolutionary stages of the SNR, or by the SNR shock expanding in a dense medium being slowed down by this medium. The first model can be applicable especially to the SE region of 3C 391 where a break-out morphology is evident from the radio data, while the later one is more suitable to the NW region of 3C 391 where the cloud material gets denser and cooler.  

%For X-rays
We also studied the plasma structure of MM SNR 3C 391 using {\it Suzaku} XIS data. The X-ray spectra of the SNR are well represented with the VNEI plus RRC model. We discovered RRC of H-like Si and S at  $\sim$ 2.7 and $\sim$ 3.5 keV. We discussed the possible electron cooling mechanisms for 3C 391 and we concluded that both thermal conduction and adiabatic cooling scenarios are possible explanations for the existence of RRC found in 3C 391. 

%%% Acknowledgements
\acknowledgments
We thank the referee for the detailed and constructive comments and suggestions on the manuscript. We are also grateful to the Fermi-LAT collaboration members for their support. Special thanks to J. Lande for his helpful comments on {\it pointlike} and to L. Tibaldo for his insightful suggestions about the Fermi-LAT analysis. T. Ergin acknowledges support from the Scientific and Technological Research Council of Turkey (T\?B\3TAK) through the B\3DEB-2232 fellowship program. E. N. Ercan thanks to Bogazici University for the financial support through the BAP project-5052.
$~$

$~$
 
%%% Facilities
Facilities: \facility{Fermi}, \facility{Suzaku}, \facility{Harvard-Smithsonian Center for Astrophysics 1.2 m MMW-radio Telescope}.
%\newpage
%%% Bibliography

%%% Tables
%\clearpage
\begin{table*}
 \begin{minipage}{170mm}
 \begin{center}
   \begin{tabular}{ccccc}
  \hline\hline
Component&Parameters & Whole& NW & SE \\
\hline
wabs&N$_{\rm{H}}$ [10$^{22}$ cm$^{-2}$]&3.1 $\pm$ 0.1&3.4 $\pm$ 0.1 &2.9 $\pm$ 0.1\\
VNEI&kT$_{e}$ [keV] &0.58 $\pm$ 0.01 & 0.61 $\pm$ 0.01 & 0.54 $\pm$ 0.01 \\
&Mg (solar)&1.2 $\pm$ 0.1 &1.6 $\pm$ 0.2 &1.4 $\pm$ 0.1\\
&Si (solar)&0.9 $\pm$ 0.1 &0.9 $\pm$ 0.1 &1.1 $\pm$ 0.1 \\
&S (solar) &0.8 $\pm$ 0.1 &0.7 $\pm$ 0.1 &0.8 $\pm$ 0.1 \\
&$\tau$ [10$^{12}$ cm$^{-3}$ s] &1.8 $\pm$ 0.2 &1.7 $\pm$ 0.3 &1.8 $\pm$ 0.1 \\
&Norm [ph cm$^{-2}$ s$^{-1}$]&4.1 $\pm$ 0.4 &3.1 $\pm$ 0.5 &3.2 $\pm$ 0.9 \\
\hline
Al K$\alpha$ &E [keV] & 1.58 (fixed)& 1.58 (fixed)& 1.58 (fixed)\\
&$\sigma$ (keV)& 0 (fixed)& 0 (fixed)& 0 (fixed)\\
&Norm [10$^{-4}$ph cm$^{-2}$ s$^{-1}$]& 2.1 $\pm$ 0.2 &2.2 $\pm$ 0.4& 2.1 $\pm$ 0.1\\
Si Ly$\alpha$ &E [keV] & 2.0 (fixed)& 2.0 (fixed)& 2.0 (fixed)\\
&$\sigma$ (keV)& 0 (fixed)& 0 (fixed)& 0 (fixed)\\
&Norm [10$^{-4}$ph cm$^{-2}$ s$^{-1}$]& 4.9 $\pm$ 0.5 &3.8 $\pm$ 0.3& 3.9 $\pm$ 0.2\\
S Ly$\alpha$ & E [keV]& 2.6 (fixed)& 2.6 (fixed)& 2.6 (fixed)\\
&$\sigma$ (keV)& 0 (fixed)& 0 (fixed)& 0 (fixed)\\
&Norm [10$^{-4}$ph cm$^{-2}$ s$^{-1}$]& 3.6 $\pm$ 0.2 & 3.1 $\pm$ 0.3& 2.9 $\pm$ 0.2\\
\hline
RRC H-like Si & E [keV] & 2.666 (fixed)& 2.666 (fixed)&2.666 (fixed)\\
&Norm [10$^{-4}$ph cm$^{-2}$ s$^{-1}$]& 5.2 $\pm$ 0.3 & 4.3 $\pm$ 0.6 & 4.1 $\pm$ 0.3\\
RRC H-like S &E [keV]& 3.482 (fixed)& 3.482 (fixed)& 3.482 (fixed)\\
&Norm [10$^{-4}$ph cm$^{-2}$ s$^{-1}$] & 4.4 $\pm$ 0.5 & 3.7 $\pm$ 0.4 &3.6 $\pm$ 0.2\\
\hline
&$\chi^2$/d.o.f &860/717=1.2 &452/361.6=1.25 & 301/251=1.2 \\
\hline
\end{tabular}
 \caption{\small{Best-fit spectral parameters of 3C 391 with corresponding errors at the 90\% confidence level in the 1.0 $-$ 5.0 keV band for an absorbed VNEI and RRC models for three regions shown in Figure \ref{figure_5a}. }}
  \label{table_2}
  \end{center}
\end{minipage}
\end{table*}

%%% End of Document
\end{document}